\documentstyle[11pt,epsf]{article}
\begin{document}
\centerline{\bf RELATIVE PROPERTIES OF SMOOTH TERMINATING BANDS}
\vspace{0.2cm}
\centerline{A. V. Afanasjev$^{1,2}$ and I. Ragnarsson$^1$}
\vspace{0.2cm}
\centerline{$^1$Department of Mathematical Physics, 
Lund Institute of Technology}
\centerline{PO Box 118, S-22100, Lund, Sweden}
\centerline{$^2$ Nuclear Research Center, Latvian Academy of Sciences}
\centerline{LV-2169, Salaspils, Miera str. 31, Latvia}
\vspace{0.5cm}
\setlength{\baselineskip}{0.4cm}

\begin{minipage}[h]{14.8cm}
\setlength{\baselineskip}{0.2cm}

{\bf Abstract.} {\it
The relative properties of smooth terminating bands observed 
in the $A\sim 110$ mass region are studied 
within the effective alignment approach. Theoretical
values of $i_{eff}$ are calculated using the configuration-dependent 
shell-correction model with the cranked Nilsson
potential. Reasonable agreement with experiment shows that
previous interpretations of these bands
are consistent with the present study.
Contrary to the case of superdeformed bands,
the effective alignments of these bands 
deviate significantly from the pure single-particle 
alignments $\langle j_x \rangle$ of the corresponding
orbitals. This indicates that in the case of smooth 
terminating bands, the effects associated with changes 
in equilibrium deformations contribute significantly 
to the effective alignment.}
\end{minipage}
\vspace{0.5cm}

\section{Introduction}

With the new arrays of $\gamma$-detectors, systematic 
investigations of rapidly rotating nuclei at the limit 
of angular momentum within specific configurations 
have become possible. The corresponding rotational 
bands are referred to as {\it terminating}. These bands, 
which are collective at low spin, gradually lose their 
collectivity and
exhaust their angular momentum content approaching a
pure single-particle ({\it terminating}) state of
maximum spin $I_{max}$. The existence of a maximum spin for a
specific configuration is a manifestation of the
finiteness of the nuclear many-fermion system where
the angular momentum in the terminating state is built 
from the contributions of the valence particles and valence
holes \cite{BR.83,RXBR.86,A110}. The state of maximum spin has
oblate $(\gamma=60^{\circ})$ or prolate $(\gamma=-120^{\circ})$
non-collective shape. Keeping in mind that the termination
of a rotational band takes place at  high spin where the pairing
correlations are of small importance, its configuration can be
defined by the number of particles (holes) in different $j$-shells.
Since nucleons are fermions which obey the Pauli principle,
one valence particle in a $j$-shell contributes with $j\hbar$,
the next with $(j-1)\hbar$ etc.

In most cases this gradual interplay between
collective and single-particle degrees of freedom is
difficult to study in experiment. One typical situation,
existing for example in the $A\sim 158$ mass region
(see Ref. \cite{158Er} and references therein), is that although
the terminating states are seen in experiment, the rotational
bands from which they originate go away from the yrast line
with decreasing spin and are thus difficult to observe over
a large spin range. Another even more common situation is
that the rotational bands, which are yrast in some spin range,
end up in terminating states residing well above the yrast
line. Only recent findings of {\it smooth terminating bands}
in the $A\sim 110$ mass region \cite{Sb109,Rag95} observed
over a wide spin range have opened the possibility for a detailed
and systematic investigation of the terminating band phenomenon. 
Considering the spin range over which these bands are observed 
and the fact that in many cases
their terminating states have been definitely or tentatively
seen in experiment, this mass region indeed represents a
unique laboratory for a theoretical and experimental study of 
the gradual interplay between collective and single-particle 
degrees of freedom within the nuclear system.

One should note, however, that in many cases, as indicated in 
Table 1, the observed bands in this region are either not 
linked to the low spin level scheme or their spins and/or
parities are not well established in experiment. As a 
consequence, the interpretation that they are observed up 
to termination is partly based on comparisons with the 
$(E-E_{RLD})$ curves
for yrast and near-yrast configurations obtained in model
calculations. In other cases when such bands have been
linked to the level scheme, as for example the $^{108}$Sn(2) and
$^{110}$Sb(1) bands, the transitions depopulating the
terminating states have been established only tentatively,
see Table 1. To some extent, this reminds us of the situation 
which exists at superdeformation where only few 
superdeformed rotational bands in the $A\sim 190$ mass 
region have been linked to the low-spin level scheme 
\cite{Hg194,Pb194}. 
In such a situation, the relative properties of SD bands have 
played an important role in our understanding of their structure 
\cite{BRA.88,Rag91,Rag93}.
Similarly to the case of superdeformation, the relative 
properties of unlinked and linked smooth terminating bands 
can be very useful in order to establish if their present 
interpretation is consistent or not. In the present article,
for the first time, the relative properties of smooth terminating 
bands observed in $^{108,109}$Sn and $^{109,110}$Sb nuclei are 
studied. This is done within the configuration-dependent 
shell-correction 
approach with the cranked Nilsson potential \cite{BR.85} using the 
effective alignments of the observed bands. Preliminary results of
this investigation have been presented at the conference
in Thessaloniki, Greece \cite{Greece}.

\begin{table}[ht]
\caption{Smooth terminating bands observed in $^{108,109}$Sn 
and $^{109,110}$Sb. The experimental data are taken from 
Refs. \protect\cite{Sn108,Sn109,Sb109new,Sb110}.
Minimum $I_{min}^{exp}$ and maximum $I_{max}^{exp}$
spins observed within the bands, as measured or estimated, 
are shown in columns 
2 and 3. Note that at high spins, the statistics in 
experiment were not high enough to determine the 
multipolarities of the transitions within the bands,
which were then assumed to have stretched E2-character.
The configurations assigned to these bands are given
in Figs. 2, 3 and 4. The corresponding calculated maximum 
spin, $I_{max}^{th}$, is shown in column 4. Column 5 
shows if the observed band structures are linked to the 
low-spin level scheme or not. Symbol F  (firm) 
is used for the cases when the linking transitions 
allowed firm establishment of spins and parities 
of observed band structures, symbol N (no) for the cases
when no linking has been established
in experiment and symbol T (tentative) for the 
cases when linking transitions have been seen in 
experiment but unsufficient experimental 
information did not allow to establish firmly
either spin or parity of observed band.
Column 6 shows the energies  
$E_{\gamma}^{term}$ of $\gamma$-transitions 
depopulating terminating
states. In the cases, when they are established
only tentatively, parantheses are used.
}
\vspace{0.5cm}
\begin{center}
\begin{tabular}{|c|c|c|c|c|c|} \hline
  Band  & $I_{min}^{exp}$ &  $I_{max}^{exp}$ & $I_{max}^{th}$ & 
 Linking &  $E_{\gamma}^{term}$ \\ \hline 
1 & 2 & 3 & 4 & 5 & 6 \\ \hline
$^{108}$Sn(1)$_{bot}$$^a$ &  $12^+$  & $(32^+)$  & $36^+$      & F 
&    \\ 
$^{108}$Sn(2)       &  $(15)^b$  & $(39^-)$   & $39^-$     & F 
&  (2569) \\ 
$^{108}$Sn(3)       &  $(20^-)^c$  & $(36^-)^c$ & $38^-$   & N 
&    \\ 
$^{109}$Sn(2)       &  $(51/2^-)$  & $(79/2^-)$ & $79/2^-$ & T 
&  (2526) \\ 
$^{109}$Sn(3b)      &  $(49/2^+)$  & $(77/2^+)$ & $85/2^+$  & T 
&    \\ 
$^{109}$Sn(4)       &  $27/2^{(+)}$  & $(67/2^+)$ & $71/2^+$ & T 
&    \\ 
$^{109}$Sn(5)       &  $(37/2^-)^d$  & $(73/2^-)^d$ & 
$81/2^-(77/2^-)^e$ & T &    \\ 
$^{109}$Sb(1)       &  $(35/2^-)^c$  & $(83/2^-)^c$ & $83/2^-$ & N 
&  2737 \\ 
$^{109}$Sb(2)       &  $(43/2^+)^c$  & $(87/2^+)^c$ & $87/2^+$ & N 
&  2822 \\ 
$^{109}$Sb(3)       &  $(49/2^+)^c$  & $(89/2^+)^c$ & $89/2^+$     
& N &  2732 \\ 
$^{109}$Sb(4)       &  $(59/2^+)^c$  & $(87/2^+)^c$ & $87/2^+$     
& N &  2810 \\ 
$^{110}$Sb(1)       &  $15^{(+)}$  & $(45^+)$   &  $45^+$       & T 
&  (2762) \\ 
$^{110}$Sb(2)       &  $(19^-)^c$    & $(37^-)^c$   & $43^-(41^-)^e$ 
& N &    \\ 
$^{110}$Sb(3)       &  $(20^-)^c$    & $(38^-)^c$   & $42^-$     & N 
&     \\ 
\hline
\end{tabular}
\end{center}
\vspace{0.5cm}
Comments:
\\
$^a$ Band 1 in $^{108}$Sn consists of two branches which
cross at $I=(32^+)$. The top branch consists of only three 
transitions and the energy of lowest transition seems 
to be affected by interaction between the $(32^+)$ 
states of the bottom and top branches. Because of this
an analysis of top branch based on $i_{eff}$ cannot
be fully conclusive. As a consequence, only the bottom
branch of this band marked as $^{108}$Sn(1)$_{bot}$
throughout the manuscript is used in present study.
\\
$^b$ The states in the spin range $I=17^--33^-$ have 
firm experimental spin-parity assignment
\\
$^c$ Tentative spins and parities have been 
assigned based on comparison with  $(E-E_{RLD})$
curves for yrast and near-yrast configurations
obtained in model calculations and also based on where
the band is experimentally observed to feed into
low-spin structures.
\\
$^d$ Parity $\pi=-$ and signature $\alpha=+1/2$
are used for this band based on best agreement between
theory and experiment. Note, however, that from 
experimental point of view ($\pi=+$, $\alpha=-1/2$) 
is an alternative possibility, see Ref. \protect\cite{Sn109} 
for detail.
\\
$^e$ Maximum spin of configuration as defined
from the distribution of particles and holes 
at low spin is shown in parantheses. The fact 
that calculated values of the spin of terminating 
state is $2\hbar$ higher indicates that close to 
termination, the highest in energy 
$\nu (g_{7/2}d_{5/2})$ particle ``jumps''
to the lowest $\nu (d_{3/2}s_{1/2})$ orbital,
see Fig. 7 and text for details.
\end{table}

The article is organized in the following way. In section 2,
we outline in brief the configuration-dependent shell correction
model and the special features used in our calculations.
In addition, the effective alignment approach \cite{Rag91}
is analysed for the case of smooth terminating bands.
In section 3, the calculated effective alignments of the
bands, which differ in occupation of one orbital, are compared
with experiment. Finally, section 4 summarizes our main 
conclusions.

\section{Theoretical method}

So far, the smooth terminating band phenomenon 
\cite{A110,Sb109,Rag95,Sn106,Sn108,Sn109,Sb109new,Sb110,Te114,Te116,I113a,I115b}
has been studied in detail only with the cranked Nilsson 
potential using the 
configuration-dependent shell-correction approach 
\cite{A110,BR.85}. Since 
the details of this approach can be found in Refs. 
\cite{A110,BR.85}, only 
the main features will be outlined
briefly below. The cranking Hamiltonian
is diagonalized in a rotating harmonic oscillator basis. In order
to facilitate the identification of the different orbitals,
the small couplings between the different $N$-shells 
($N_{rot}$-shells) are neglected so that each orbital 
definitely belongs to one $N$-shell. Furthemore, 
virtual crossings between the single-particle orbitals 
within the $N$-shells are removed in an approximate way 
and, as a result, smooth diabatic orbitals are obtained. 
An additional feature \cite{A110,Rag95} is that the 
high-$j$ orbitals in each $N$-shell are identified after 
the diagonalization. As a result, it is possible to
distinguish between the particles of (approximate) 
$g_{9/2}$ character and the particles belonging to other
subshells of the $N=4$ shell. This feature is
crucial in the specification of the number of particles
excited across the $Z=50$ spherical shell gap; a 
situation typical in smooth terminating bands of the 
$A\sim 110$ mass region. Only with this specific feature
implemented it is possible to trace fixed configurations 
as a function of spin all the way up to termination.
At a typical deformation of 
$\varepsilon_2\sim 0.2-0.3$, the low-$j$ subshells are 
rather strongly mixed. For example, in the $N=4$ shell
the subshells $g_{7/2}$, $d_{5/2}$, $d_{3/2}$ and 
$s_{1/2}$ are treated as one entity in our formalism.
The corresponding Nilsson orbitals can however generally
be classified as having their main components in either the
($g_{7/2}d_{5/2}$) subshells or the ($d_{3/2}s_{1/2}$) 
subshells and we will often refer to configurations as
having a fixed number of particles in these subgroups. 
The calculations are carried out in 
a mesh in the deformation space, 
$(\varepsilon_2,\varepsilon_4,\gamma)$.
Then for each fixed configuration and each spin 
separately, the total energy of a nucleus is determined 
by a minimization in the shape degrees of freedom.
Pairing correlations are neglected so the calculations
can be considered fully realistic for spins above
$I \sim 20\hbar$ in the $A \sim 110$ mass region. An additional 
source of possible discrepancies between experiment and 
calculations at low spin could be the one-dimensional 
cranking approximation used in the present approach. 

Effective alignment of two bands (A and B) is simply the 
difference between their spins at constant 
$\gamma$-transition energy $E_{\gamma}$
(or rotational frequency $\hbar\omega$) \cite{Rag91}: 
\begin{equation}
i^{B,A}_{eff}(E_{\gamma})=I^B(E_{\gamma})-I^A(E_{\gamma})
\end{equation}
which is illustrated in Fig. 1. Experimentally, 
$i_{eff}$ includes both the alignment of the single-particle 
orbital and the effects associated with changes in deformation, 
pairing etc. between two bands. This approach exploits the fact
that spin is quantized, integer for even nuclei and
half-integer for odd nuclei and furthermore constrained
by signature. One should note that with the
configurations and specifically the signatures
fixed, the relative spins of observed bands
can only change in steps of $\pm 2\hbar \cdot n$
($n$ is an integer number).

\input epsf
\vspace*{0.5cm}
        \epsfysize 12.0cm  \epsfbox{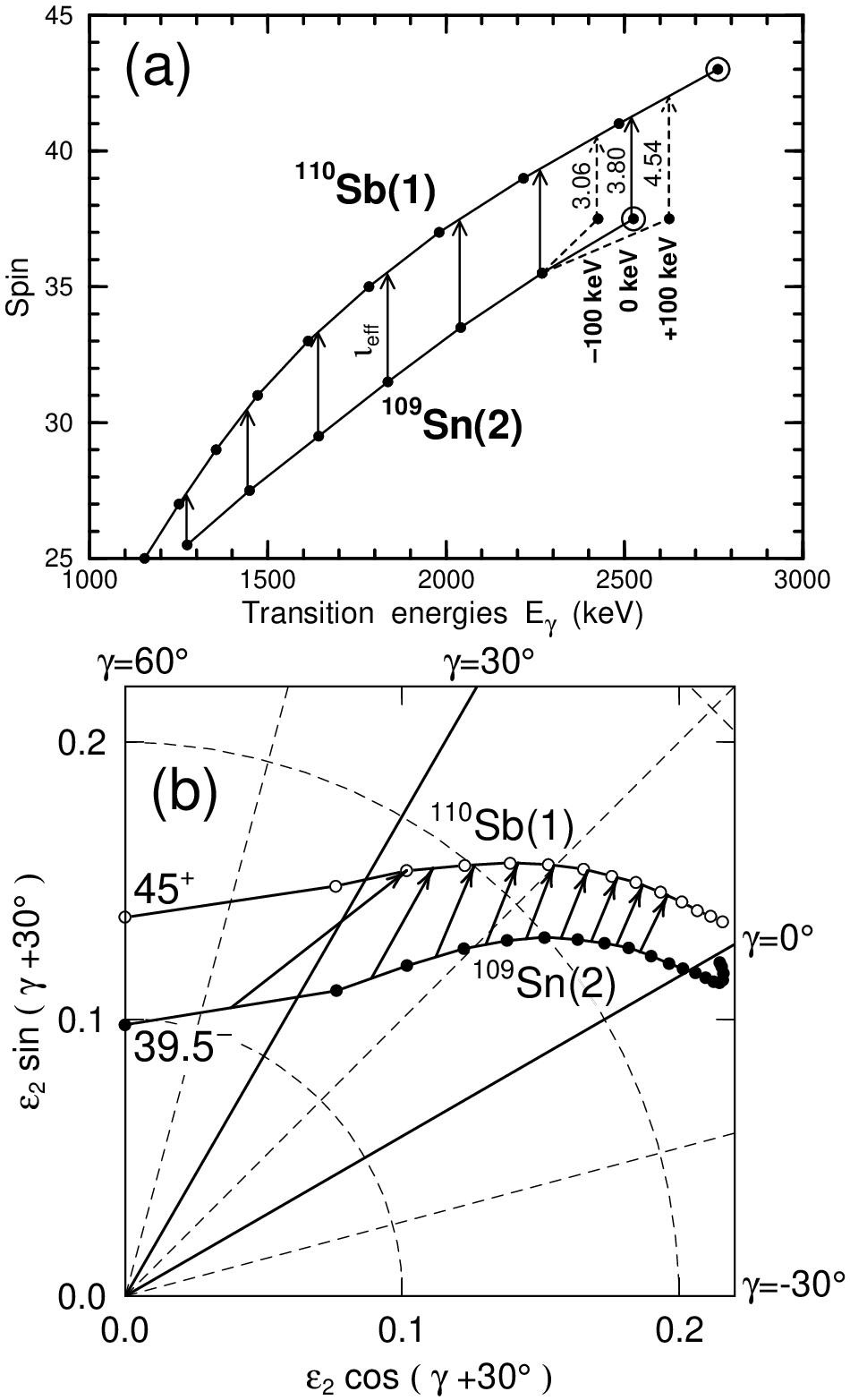}
\vspace*{0.0cm}
Fig. 1. {\it (a) Spin vs. transition energy for the observed 
$^{109}$Sn(2) and $^{110}$Sb(1) bands with the highest energy 
transitions (which depopulate the terminating states)
encircled.  
The arrows show the effective alignment $i_{eff}$ in the 
$^{109}$Sn(2)/$^{110}$Sb(1) pair calculated at the 
energies of $\gamma$-transitions in the $^{109}$Sn(2) 
band. Dashed lines (arrows) illustrate the 
situation when the energy of $\gamma$-transition 
depopulating the terminating state in $^{109}$Sn(2) band
is changed by $\pm 100$ keV which is only $\approx 4\%$
of total energy of this transition. Corresponding 
changes in effective alignment $i_{eff}$ (in units 
$\hbar$) are indicated. (b) Calculated shape trajectories in 
the $(\varepsilon_2,\,\gamma)$ plane for configurations
assigned to the $^{109}$Sn(2) and $^{110}$Sb(1) bands.
The deformation points are given is steps of $2\hbar$.
The spin values for terminating states of these 
configurations are given. The arrows
are used to indicate the difference in deformation
between the two bands at rotational frequencies where 
the theoretical effective alignment $i_{eff}$ is 
extracted (Fig. 6c below).
Note that these deformations are obtained from 
the interpolations in the $(\varepsilon_2,\gamma)$
plane. In the reference 
band ($^{109}$Sn(2)), these values for the 
transitions depopulating the states with spin $I$
are exactly equal to the average deformation of 
the $I$ and $(I-2)$ states.}
\vspace{0.5cm}

The effective alignment approach has been succesfully 
applied for an interpretation of the structure of SD 
bands observed in the $A\sim 140-150$ mass region 
employing present model \cite{Rag91,Rag93,Gd,Gd147}
and cranked relativistic 
mean field theory \cite{ALR.97}. One should note
that compared with the case of SD bands there 
are several essential differences in the case
of smooth terminating bands. Most of SD bands 
are near-prolate with rather small $\gamma$-deformation. 
In addition, the changes in equilibrium deformation 
with increasing rotational frequency are rather similar 
in the SD bands of neighbouring nuclei. This indicates
that the difference in equilibrium deformation between 
two SD bands under comparison stays rather constant when 
the rotational frequency is varied. 
As a result, the effective alignment $i_{eff}$ 
of two SD bands is predominantly defined by 
alignment properties of single-particle orbital 
by which these bands differ \cite{Rag91}. 

Contrary to this, the smooth terminating bands
in the $A\sim 110$ region undergo drastic shape 
changes from near-prolate at low spin to oblate 
for the terminating state, see Fig. 1b. 
Moreover, as indicated by arrows
in Fig. 1b, the difference in equilibrium 
deformation between compared bands at rotational 
frequencies, where the values of $i_{eff}$ are 
extracted, shows considerable dependence from 
spin. It is especially pronounced for the 
transition depopulating the terminating state. 
All this suggest that in the case 
of smooth terminating bands, the shape changes 
play a more important role for the 
quantitave value of $i_{eff}$ compared with the 
case of SD bands. Fig. 1a shows that the 
effective alignment $i_{eff}$ is a 
sensitive probe of relative properties of 
smooth terminating bands. This is 
illustrated on the example of the 
$^{109}$Sn(2)/$^{110}$Sb(1) pair. Changes in 
the energy of the transition depopulating the
terminating state in $^{109}$Sn(2) band by 
$\pm 100$ keV 
(which is only $\sim 4\%$ of energy of this 
transition) lead to a change in $i_{eff}$ of
$\pm 0.74\hbar$ which is $\sim 20\%$ of
original value, $i_{eff}=3.8\hbar$. Note
that the transition depopulating the 
terminating state links two states having the 
largest difference in equilibrium deformation 
within the band: the terminating state has 
$\gamma=60^{\circ}$ while the state with 
($I_{max}-2$) has $\gamma\sim 30^{\circ}$
and a larger $\varepsilon_2$ than the 
terminating state. As a result, the accuracy 
in describing $i_{eff}$ for the transition 
depopulating the terminating state can be a very
accurate measure of how well theory describes
the shape changes close to termination, 
induced by the particle in the active 
orbital. 

\section{Results and discussion}

In the present study we consider all bands 
in $^{108,109}$Sn and $^{109,110}$Sb built
on $2p-2h$ excitations across the spherical
$Z=50$ shell gap for which configurations
have been suggested in the literature
\cite{Sn108,Sn109,Sb109new,Sb110}. It is
in these nuclei that bands which are 
interpreted as going to termination
(definitely or tentatively) are observed,
namely,  the $^{108}$Sn(2), $^{109}$Sn(2), 
$^{109}$Sb(1-4) and $^{110}$Sb(1) bands.
Some properties of  these bands are
collected in Table 1. The pairs 
of bands which differ by the occupation 
of one orbital are indicated in Fig. 2.
The shorthand notation $[p_1p_2,n]^{\alpha_{tot}}$
\cite{A110} is used for configuration labelling.
In this notation $p_1$ is the number of $g_{9/2}$ 
proton holes, $p_2$ the number of $h_{11/2}$ protons, 
$n$ the number of $h_{11/2}$ neutrons and 
$\alpha_{tot}$ is the total signature of the 
configuration. Since the signature depends on the type 
of nucleus (even or odd),
the following simplification is used:
($\alpha_{tot}$=+)$\equiv$($\alpha_{tot}$=+1/2(odd))
$\equiv$($\alpha_{tot}$=0(even));\,\,\,
($\alpha_{tot}$=--)$\equiv$($\alpha_{tot}$=--1/2(odd))
$\equiv$($\alpha_{tot}$=+1(even)).
Note the similarity of this
shorthand notation to the one used for the
SD bands. In both cases, the configuration
is specified by the number of high-$j$ 
intruder orbitals occupied. In addition, the 
number of emptied high-$j$ extruder orbitals 
is also specified in the case of smooth 
terminating bands. Such configuration notation, 
being extremely useful and simple, does also
provide partial information about 
non-intruder orbitals occupied. 
In the $A\sim 110$ mass region, these 
non-intruder orbitals belong to the proton 
and neutron $(g_{7/2} d_{5/2})$ subshells
(with some admixture from $d_{3/2}$ and $s_{1/2}$).
The detailed structure of the configurations assigned 
to the bands is given in Figs. 3 and 4.
In the present study, we use the same configuration 
and spin assignments for the observed bands as 
in Refs. \cite{Sn108,Sn109,Sb109new,Sb110}. The 
comparison between experiment and calculations is 
presented in Figs. 5, 6.

\subsection{Pairing correlations and accidental
degeneracies}

At low rotational frequencies $\hbar \omega \leq 0.6-0.7$ MeV, 
the observed bands are influenced by pairing 
correlations which in many cases lead to 
paired band crossings. This is seen in 
Figs. 5, 6 as characteristic sharp changes 
in $i_{eff}$, which appear if at least one of
the compared bands undergoes a paired band crossing.
The experimental points which  are influenced
by pairing correlations, as follows e.g. from
the analysis of dynamic moments of inertia
$J^{(2)}$, are shown on shaded
background in Figs. 5, 6. 
In general, these points should not be taken
seriously into account when theoretical
results are compared with experiment since
pairing correlations are neglected in 
the calculations. Note, however, that in some 
cases, for example in the 
$^{109}$Sb(1)/$^{110}$Sb(1) pair in 
the energy range $\sim 1100-1400$ keV
and in the  $^{109}$Sb(3)/$^{110}$Sb(1) pair 
in the energy range $\sim 1200-1700$ keV,
see Figs. 5a and 5h,
the results of calculations are reasonably
close to experimental data even in the 
rotational frequency range where the pairing 
correlations are expected to play a role.  
This suggests that the occupation of the orbital,
by which two bands differ, does not introduce
considerable changes in the pairing field.

\vspace*{1.7cm}
        \epsfxsize 14.0cm  \epsfbox{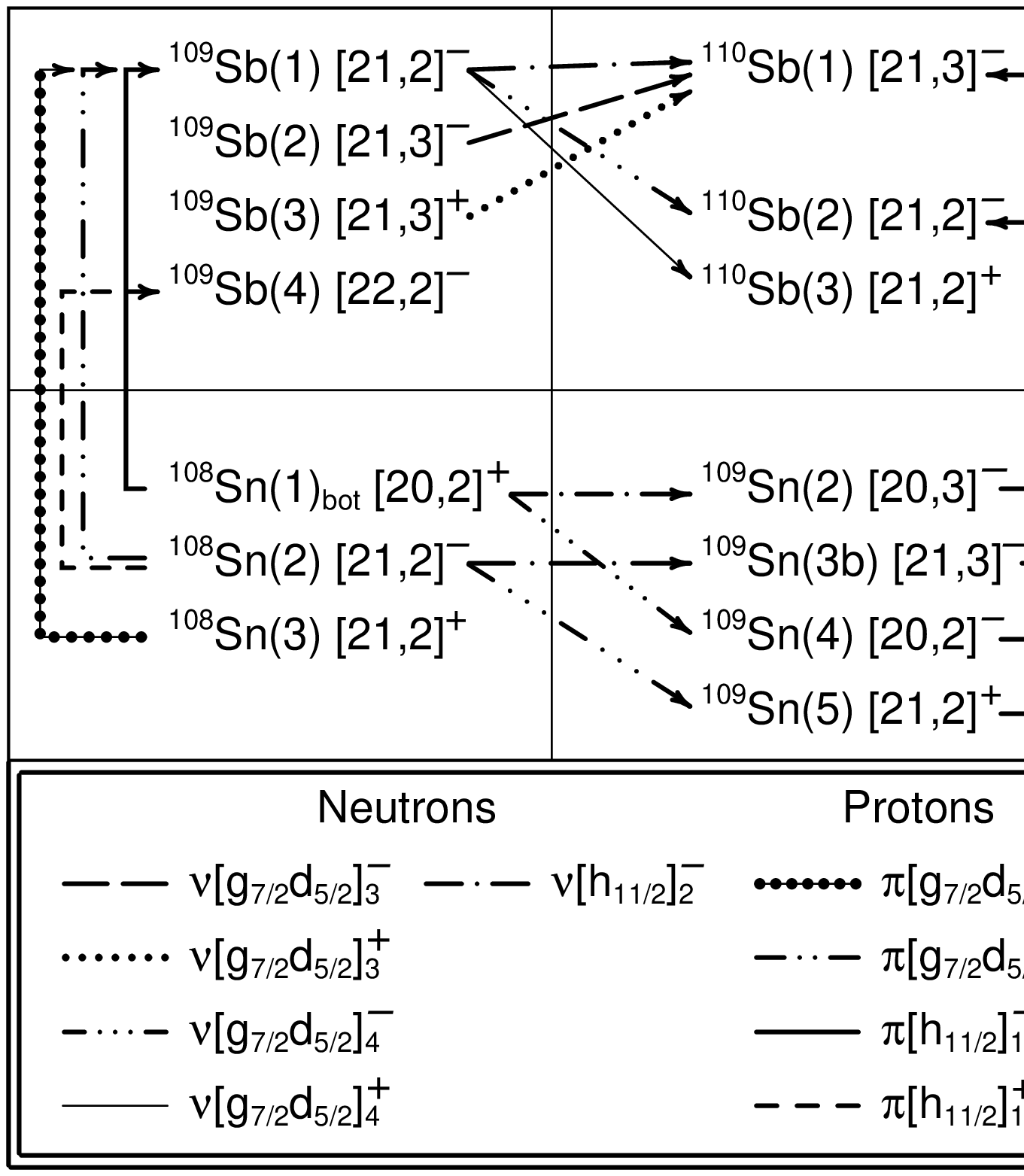}
\vspace*{-1.8cm}
Fig. 2. {\it Smooth terminating bands observed in $^{109,110}$Sb
and in $^{108,109}$Sn \protect\cite{Sn108,Sn109,Sb109new,Sb110}. 
The assigned configurations are indicated after band
labels. The difference in the configurations of
observed bands related to specific orbitals is indicated
by arrows of different types. The correspondence between
the type of arrow and orbital is given in bottom
panel. The orbitals are labeled by 
the dominant components of their wave functions at 
$\hbar \omega=0.0$ MeV, by the sign of their signature 
$\alpha$ given as superscript and by the position of 
the orbitals within the specific signature group given as 
subscript.} 
\vspace{0.5cm}

In addition, some unsmoothness of experimental 
$i_{eff}$ values seen at low and sometimes at
medium spins could arise from accidental closeness 
in energy of two states with the same quantum 
numbers belonging to different bands. For example, 
the energy of the bottom transition in band 2 of 
$^{109}$Sn seems to be disturbed because of 
interaction between the $(51/2^-)$ states of bands 
1 and 2 \cite{Sn109}. Similarly, the two $32^+$
states belonging to the two branches of band 1
in $^{108}$Sn \cite{Sn108} come very close together
and are consequently shifted in energy.
These disturbancies in 
$i_{eff}$ are seen in the $^{108}$Sn(1)$_{bot}$/$^{109}$Sn(2)
and $^{108}$Sn(1)$_{bot}$/$^{109}$Sb(1) pairs,
see Figs. 5c and 6a.

\subsection{General features of $i_{eff}$ for frequencies
$\hbar \omega \geq 0.6-0.7$ MeV}

At higher rotational frequencies $\hbar \omega \geq 0.6-0.7$ MeV,
the lack of expected quasiparticle alignments in 
the observed bands of the $A\sim 110$ mass region 
has been attributed to a loss of static pairing 
correlations \cite{A110,Sb109,Sn106,LF.94,RW.93}. Indeed,
the experimental dynamic moments of inertia 
$J^{(2)}$ of the most of  observed bands is
smoothly decreasing at these frequencies 
which suggest that our calculations should
be able to describe the experiment at these 
frequencies. If we do not consider the 
effective alignment for transitions 
depopulating terminating states which are
discussed below, then in most cases there
is a reasonable agreement between calculations 
and experiment, see Figs. 5 and 
6. Only for the cases of the 
$^{108}$Sn(3)/$^{109}$Sb(1) and 
$^{108}$Sn(1)$_{bot}$/$^{109}$Sn(4) 
pairs the disagreement with experiment 
is considerable, see Figs. 5f and 6h. 
For the comparisons where the $\nu [g_{7/2} d_{5/2}]^-_4$
orbital is active, one notes that experimental effective 
alignment $i_{eff}$ of the 
$^{108}$Sn(1)$_{bot}$/$^{109}$Sn(4) 
pair is more downsloping with increasing
$E_{\gamma}$ than the ones seen in the 
$^{108}$Sn(2)/$^{109}$Sn(5) and
$^{109}$Sb(1)/$^{110}$Sb(2) pairs,
see Figs. 5d, 5e and 5f.
The modest discrepancies in the two latter cases are 
discussed below while we have no explanation for 
the larger discrepancies in the $^{108}$Sn(3)/$^{109}$Sb(1) and 
$^{108}$Sn(1)$_{bot}$/$^{109}$Sn(4) 
pairs. Systematic discrepancies seen
for the $^{108}$Sn(3)/$^{109}$Sb(1) pair
certainly put some doubts on the configuration 
assignment for the unconnected band 3 in $^{108}$Sn.

\vspace*{2.7cm}
        \epsfxsize 14.0cm  \epsfbox{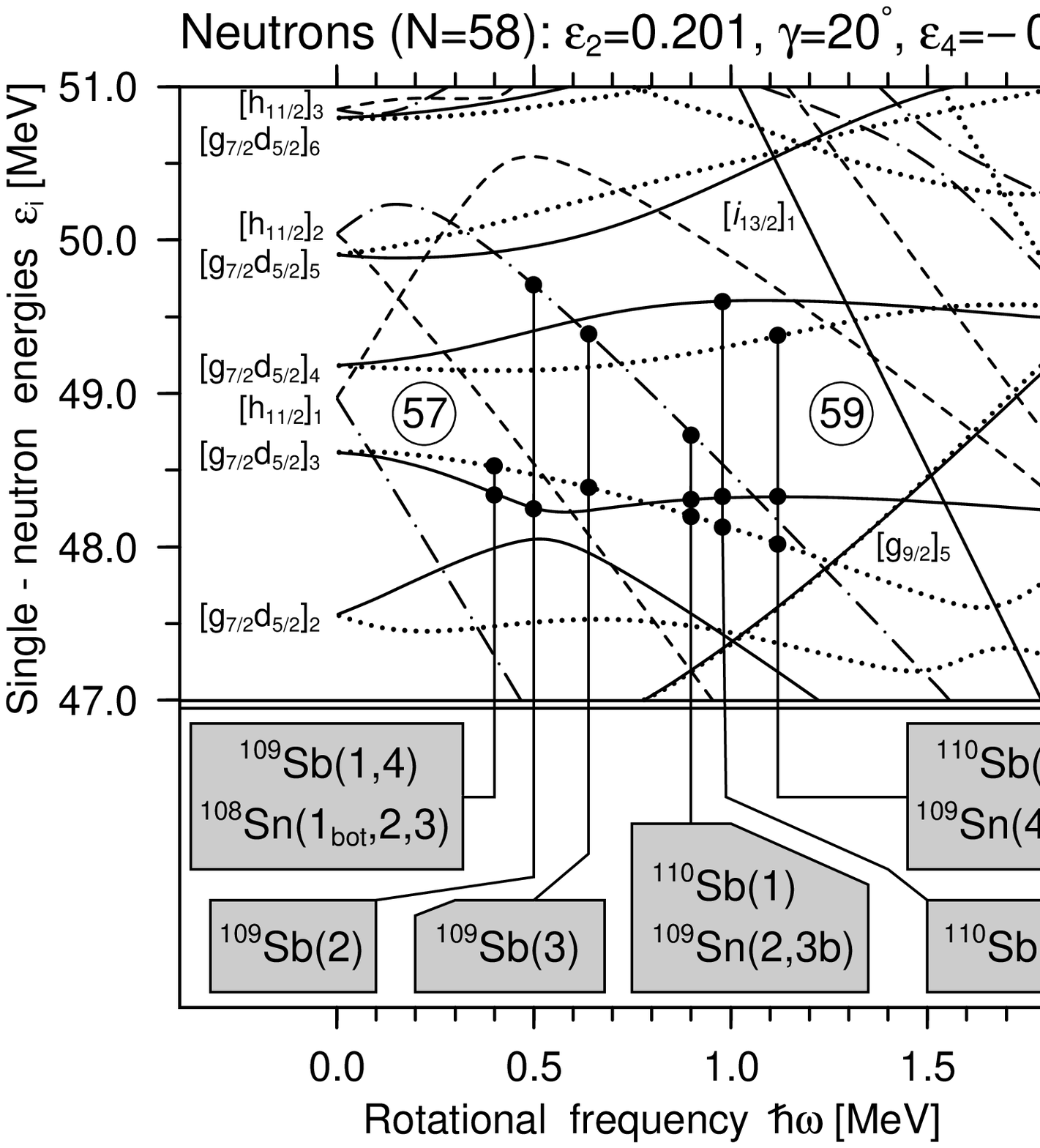}
\vspace*{-2.3cm}
Fig. 3. {\it Neutron single-particle energies
in the rotating frame (routhians) drawn at the calculated
equilibrium deformation ($\varepsilon_{2}=0.201,\,\,\gamma=20^{\circ},
\,\,\varepsilon_{4}=-0.016$) of the $I=33.5^-$ state of configuration 
assigned to the $^{109}$Sb(1) band. At $\hbar\omega=0$ MeV, 
the orbitals are labeled by dominant components of their
wave functions, and as a subscript the position of the 
orbitals within the specific group. The following convention
is used: $(\pi=+,\,\,\alpha=+1/2)$ solid line,
$(\pi=+,\,\,\alpha=-1/2)$ dotted line,
$(\pi=-,\,\,\alpha=+1/2)$ dashed line,
$(\pi=-,\,\,\alpha=-1/2)$ dot-dashed line, i.e. dots correspond 
to signature
$\alpha=-1/2$ and dashed lines to negative parity.
The single-particle energies and rotational frequencies are
transformed from the oscillator units used in the computer code
to physical units (MeV) according to following expressions
$E({\rm MeV})=41A^{-1/3}(1\pm \frac{N-Z}{3A}) E({\rm osc.units})$
and $\hbar \omega ({\rm MeV})= S \cdot
41A^{-1/3}(1\pm \frac{N-Z}{3A}) \hbar \omega ({\rm osc.units})$,
where + ($-$) signs are used for neutrons (protons),
respectively. In these expressions, $Z$, $N$ and $A$ are proton,
neutron and mass numbers, respectively, and $S$ is the Strutinsky
renormalization factor calculated at the selected deformation.
The occupation of different orbitals in the bands under study
is shown. Note that the $[g_{7/2}d_{5/2}]_2$ orbitals as well
as the lowest in energy $h_{11/2}$ orbital of each signature 
are occupied in all bands.}
\vspace{0.0cm}

\subsection{Admixture of the $(d_{3/2}s_{1/2})$ subshell
at highest observed frequencies}

Amongst the bands under study, only $^{109}$Sn(4,5) 
and $^{110}$Sb(2) bands show an increase in $J^{(2)}$ 
at highest observed frequencies \cite{Sn109,Sb110}. 
This increase in $J^{(2)}$ correlates with more or 
less sharp changes in effective alignments seen
at highest rotational frequencies in the pairs in 
which one of these bands is involved, see Figs. 
5d, 5e, 5f, 6b  and 6f.
With exception of the  $^{108}$Sn(1)$_{bot}$/$^{109}$Sn(4)
pair the average gain in $i_{eff}$ is $\approx 0.5\hbar$.
In the $^{108}$Sn(1)$_{bot}$/$^{109}$Sn(4) pair, 
larger changes in $i_{eff}$ for highest transition
energies are connected with disturbances in transition
energies depopulating the $(32^+)$ state of the 
$^{108}$Sn(1)$_{bot}$ band, see discussion above.
These observations indicate that contrary to other 
bands, the configurations of the $^{109}$Sn(4,5) and 
$^{110}$Sb(2) bands do not remain pure at highest 
observed frequencies. One should note that these changes 
in $i_{eff}$ are not reproduced in the calculations.
The question is then how to understand the origin
of the increase in $J^{(2)}$ of these bands and 
changes in $i_{eff}$ of the three compared pairs.
Note that neutron configurations of these three bands 
specified in terms of occupation of different signature
orbitals are the same, namely, 
$\nu [(g_{7/2} d_{5/2}) (\alpha=-1/2)]^4 
[(g_{7/2} d_{5/2}) (\alpha=+1/2)]^3 
[h_{11/2} (\alpha=-1/2)]^1 [h_{11/2} (\alpha=+1/2)]^1$,
see Fig. 3. Specifically, it is only in these bands 
that the fourth $(g_{7/2} d_{5/2}) (\alpha=-1/2)$ orbital 
is occupied which strongly suggests that the specific
properties of these bands are related to some features 
of this orbital. 

\vspace*{2.7cm}
        \epsfxsize 14.0cm  \epsfbox{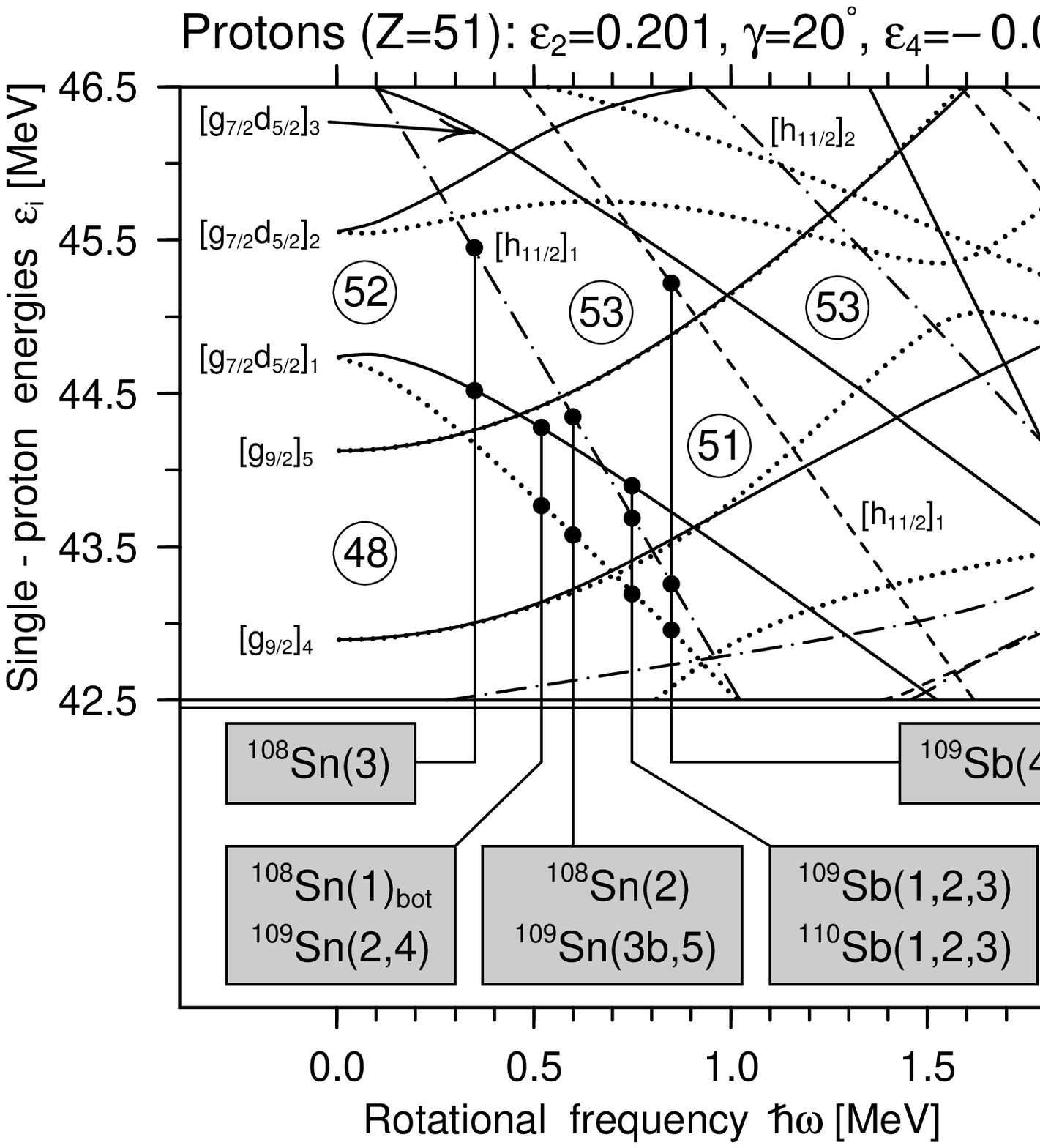}
\vspace*{-2.3cm}
Fig. 4. {\it
Same as Fig. 3, but for proton 
routhians. Note that in all configurations assigned
to the bands, the $\pi [g_{9/2}]_4$ orbitals are occupied
while the $\pi [g_{9/2}]_5$ orbitals are empty.}
\vspace{0.5cm}

It is interesting to note that terminating 
states in the configurations assigned to the $^{110}$Sb(2)
and $^{109}$Sn(5) bands are calculated at two spin units higher
than maximum spin defined from the distribution of 
the valence particles and holes over the $j$-shells at low 
spin, see Fig. 3 in Ref. \cite{Sn109} and Table 1. 
The single-neutron routhian diagrams drawn at the calculated 
equilibrium deformations of the configuration assigned to 
$^{109}$Sn(5) two and four spin units below the actual
termination are shown in Fig. 7. An analysis of these
diagrams suggests that the unpaired ``band crossing'' between
the fourth $(g_{7/2} d_{5/2}) (\alpha=-1/2)$ orbital 
and the lowest $(d_{3/2} s_{1/2}) (\alpha=-1/2)$
orbital is the most probable explanation for the 
calculated features of the $^{109}$Sn(5) and $^{110}$Sb(2)
bands. This crossing also explains why the spins 
of the terminating states in these two configurations
are two spin units higher than expected one. 
The fourth $(g_{7/2} d_{5/2}) (\alpha=-1/2)$ 
orbital has a negative contribution to the total
angular momentum at termination, namely 
$-1/2\hbar$. On the other hand, the contribution
of lowest $(d_{3/2} s_{1/2}) (\alpha=-1/2)$
orbital is positive, namely, $+3/2\hbar$. 
Therefore, if
the lowest $(d_{3/2} s_{1/2}) (\alpha=-1/2)$
orbital becomes occupied at some spin instead
of the fourth $(g_{7/2} d_{5/2}) (\alpha=-1/2)$ 
orbital, the two additional spin units contribute 
at the termination. Since the orbitals belonging to the 
$g_{7/2},\,d_{5/2},\,d_{3/2},\,s_{1/2}$ 
subshells are treated as a single entity, separate 
calculations for the configurations with either the fourth 
$(g_{7/2} d_{5/2}) (\alpha=-1/2)$ orbital
or the lowest $(d_{3/2} s_{1/2}) (\alpha=-1/2)$
orbital occupied are not straightforward. One 
should note that a similar scenario has been 
discussed for the configuration assigned to band 
1 in $^{116}$Te \cite{Te116,Piaski}.

\vspace*{1.7cm}
        \epsfxsize 14.0cm  \epsfbox{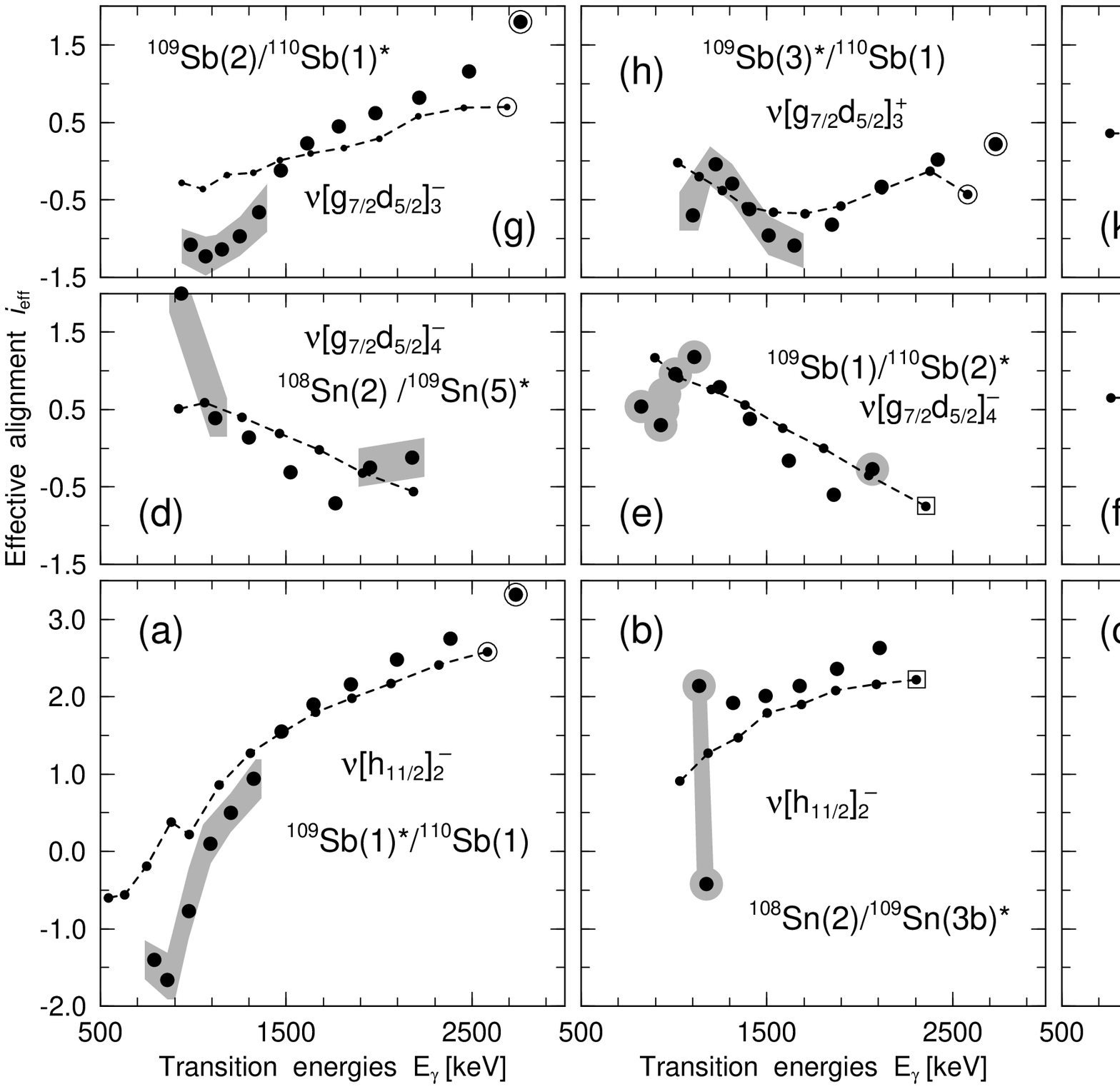}
\vspace*{-0.5cm}
Fig. 5. {\it Effective alignments associated with occupation of neutron 
orbitals, $i_{eff}$ (in units $\hbar$), extracted from experimental
bands (unconnected large symbols) are compared with the ones extracted 
from the calculated configurations assigned to them
(connected small symbols). 
The experimental effective alignment between bands A and B is 
indicated as 
``A/B''. The band A in the lighter nucleus  is taken as a 
reference so the effective alignment measures the effect of an
additional particle. The $i_{eff}$ values are shown at the 
transition energies of the shorter band indicated by an 
asterisk {\rm ($^*$)}. The compared
configurations differ by the occupation of the orbitals
indicated on the panels. The points corresponding to
transitions depopulating terminating states are
encircled. The points corresponding to transitions
depopulating the states with spin ($I_{max}-2$) are
indicated by large open squares. The experimental points 
which, as follows e.g. from the analysis of $J^{(2)}$ of the 
bands, appear to be affected either by pairing interactions 
or by unpaired band crossings are shown on shaded background.}
\vspace{0.5cm}

The fact that the changes in $i_{eff}$ of the 
pairs involving the $^{109}$Sn(4,5) and
$^{110}$Sb(2) bands are not reproduced in the
calculations, strongly suggests that the 
relative positions of the fourth  
$(g_{7/2} d_{5/2}) (\alpha=-1/2)$ orbital
and lowest $(d_{3/2} s_{1/2}) (\alpha=-1/2)$
orbital are not optimal in the present 
parametrization of the Nilsson potential. 
With the energy gap between these two orbitals
smaller by $\sim 0.5-1.0$ MeV, this crossing
will take place at a lower rotational frequency
as required by experiment. This crossing will
lead to an increase of $J^{(2)}$ as observed.
In addition, the crossing between these
two orbitals provides a qualitative explanation 
for the changes in $i_{eff}$ of three pairs 
shown in Figs. 5d, 5e and 5f.
At medium rotational frequencies, the
fourth $(g_{7/2} d_{5/2}) (\alpha=-1/2)$ orbital
is upsloping and has a negative angular momentum
alignment which is in agreement with experiment,
see Figs. 5d, 5e and 5f. In
the crossing region,  it is 
reasonable to expect an increase of 
$i_{eff}$ because the lowest 
$(d_{3/2} s_{1/2}) (\alpha=-1/2)$ orbital 
has a positive angular momentum alignment,
see Fig. 7. This is clearly seen in 
experiment, see Figs. 5d, 5e and 
5f. However, the actual gain in $i_{eff}$ 
is expected to be dependent on the 
deformation and rotational frequency at which
the crossing takes place.

\vspace*{1.0cm}
        \epsfxsize 14.0cm  \epsfbox{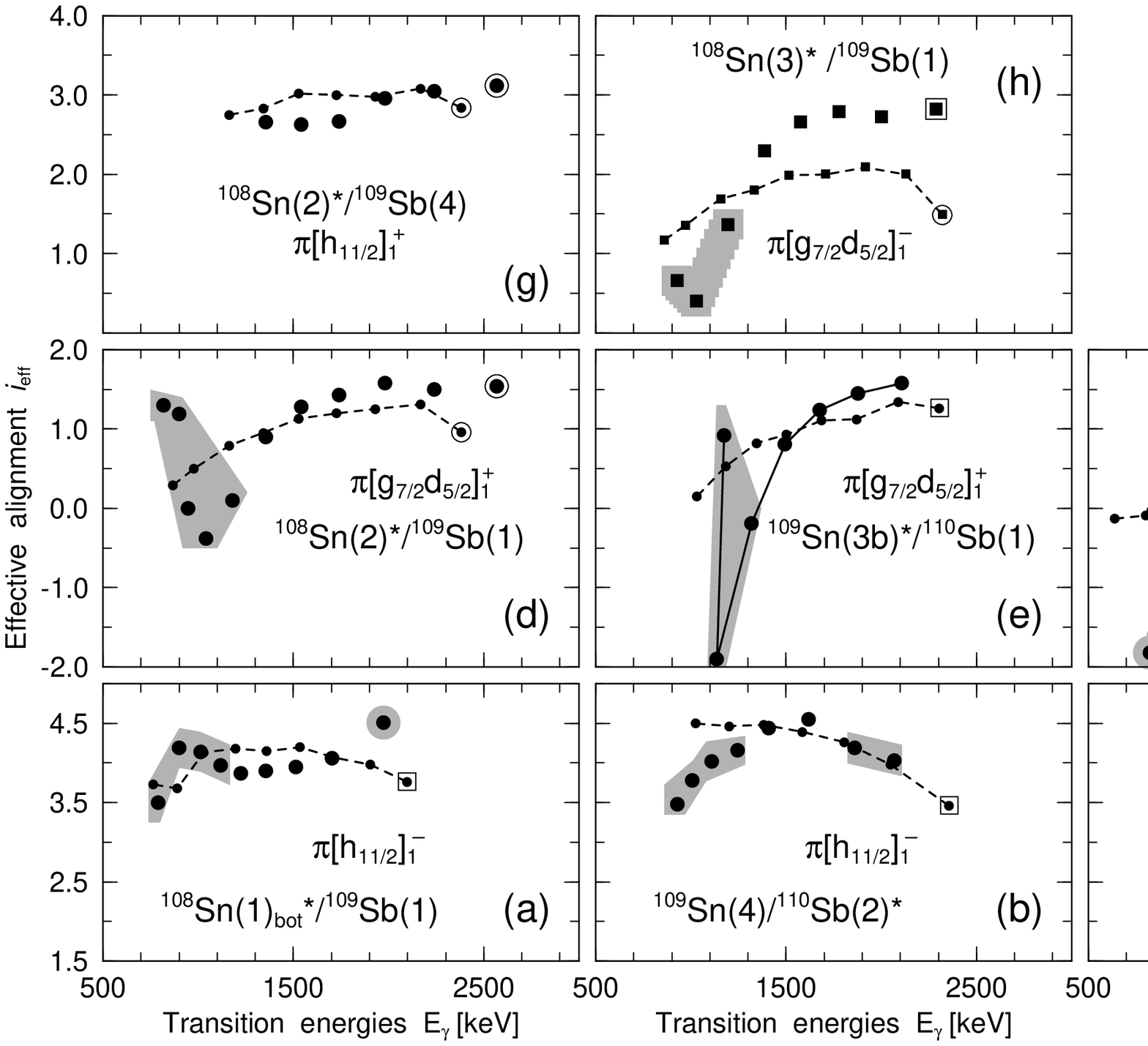}
\vspace*{-0.5cm}
Fig. 6. {\it Same as Fig. 5 but for proton
orbitals. The dotted line in the panel (c) representing the
$^{109}$Sn(2)*/$^{110}$Sb(1) pair corresponds to the 
case where the calculated energies of $\gamma$-transitions 
between the $39.5^-$ and $37.5^-$ states and between the $37.5^-$ 
and $35.5^-$ states of the configuration assigned to the 
$^{109}$Sn(2) band are increased by 166 and 33 keV, 
respectively.}
\vspace{0.5cm}

The situation discussed above appears mainly 
when neutron $(g_{7/2}d_{5/2})$ orbitals 
giving negative contributions to the total spin 
at termination are occupied. As would be expected,
they are generally  upsloping
as a function of rotational frequency, see Figs.
3 and 7, which makes the crossing with 
the downsloping lowest neutron $(d_{3/2}s_{1/2})$
orbitals possible. 
Our calculations performed for heavy Sb isotopes 
\cite{A110,ARunp} and for $^{116}$Te indicate that 
with increasing neutron number, the occupation of the 
$(d_{3/2}s_{1/2})$ 
orbitals becomes energetically favoured in many 
configurations close to termination. To some extent 
this tendency is weakened by the possibility to occupy 
the third and fourth neutron $h_{11/2}$ orbitals instead
of the $(g_{7/2}d_{5/2})$ orbitals. 

\vspace*{5.0cm}
        \epsfxsize 14.0cm  \epsfbox{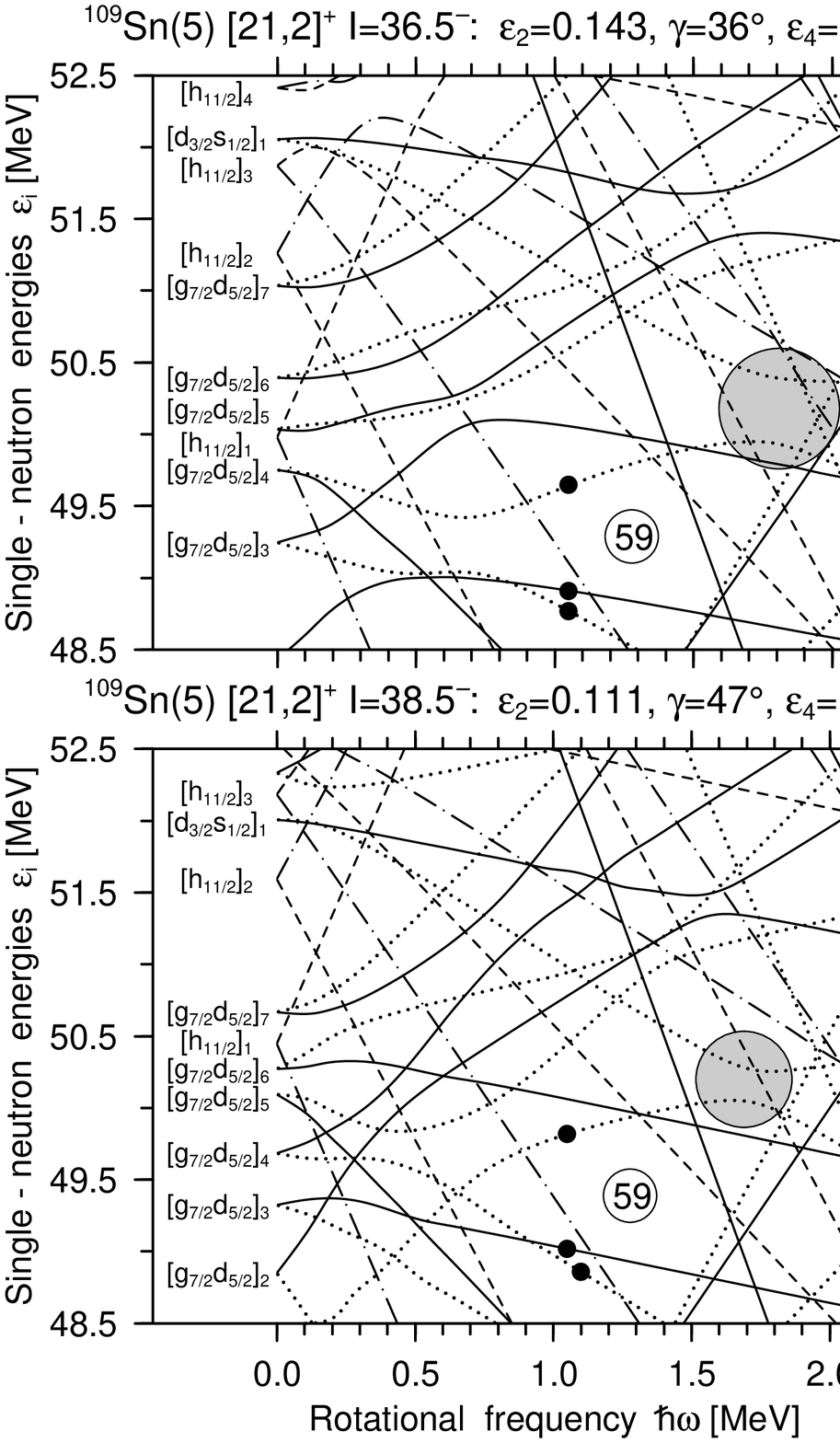}
\vspace*{0.5cm}
Fig. 7. {\it
Same as Fig. 3. 
The deformations selected correspond to equilibrium deformations
of the $I=36.5^-$ and $I=38.5^-$ states of the configuration 
$[21,2]^+$ assigned to the $^{109}$Sn(5) band. The last 
occupied $(g_{7/2} d_{5/2})$ orbitals are indicated
by solid circles. Large shaded circles are used to
outline the ``crossing'' between fourth 
$(g_{7/2} d_{5/2})(\alpha=-1/2)$ orbital and lowest
$(d_{3/2} s_{1/2})(\alpha=-1/2)$ orbital.}
\vspace{0.5cm}

\subsection{Effective alignments for transitions
depopulating terminating states}

As discussed in section 2 the effective alignment
$i_{eff}$ is a very sensitive probe of how well
the model describes the relative transition energies, 
especially for the transitions depopulating terminating states. 
Indeed, it is more difficult to reproduce the $i_{eff}$ 
values for these transitions which link the states having 
largest difference in equilibrium deformation between two 
neighbouring states within a band, see Figs. 5a,g,h, 
and 6c,d,g. In this mass region, the difference 
in $\gamma$-deformation 
(at values $\varepsilon_2\sim 0.10-0.15$)
is generally $20-30^{\circ}$.
As a result, these $i_{eff}$ values are more
sensitive both to the accuracy of reproduction
of prolate-oblate energy difference in the
liquid-drop part and the parametrization of the
Nilsson potential than the $i_{eff}$ values for
transitions at lower spin. One should note, 
however, that these discrepancies between experiment 
and calculations originate from rather small 
unadequateness in the description of transition 
energies within the bands as illustrated in Fig. 
1. Let us take as an 
example the $^{109}$Sn(2)*/$^{110}$Sb(1) pair, in 
which the largest discrepancy between experiment and 
theory is seen close to termination, see Fig. 6c. 
An excellent agreement between experiment and 
calculations shown by the dotted line in Fig. 6c
is obtained when the calculated energies of 
$\gamma$-transitions between the $39.5^-$ and $37.5^-$ 
states and between the $37.5^-$ and $35.5^-$ states of 
the configuration assigned to the $^{109}$Sn(2) band 
are increased by 166 and 33 keV, respectively. Note
that these are only few percent corrections to the
total energies of the transitions linking these states.
For the other pairs, even smaller corrections to 
the calculated transition energies are needed in order to 
get a perfect agreement between experiment and calculations 
close to termination. This clearly indicates that the 
prolate-oblate energy difference and the polarization 
effects induced by an active particle are reproduced 
reasonable well within the cranked Nilsson model for the 
bands under study.

\vspace*{1.7cm}
        \epsfxsize 18.0cm  \epsfbox{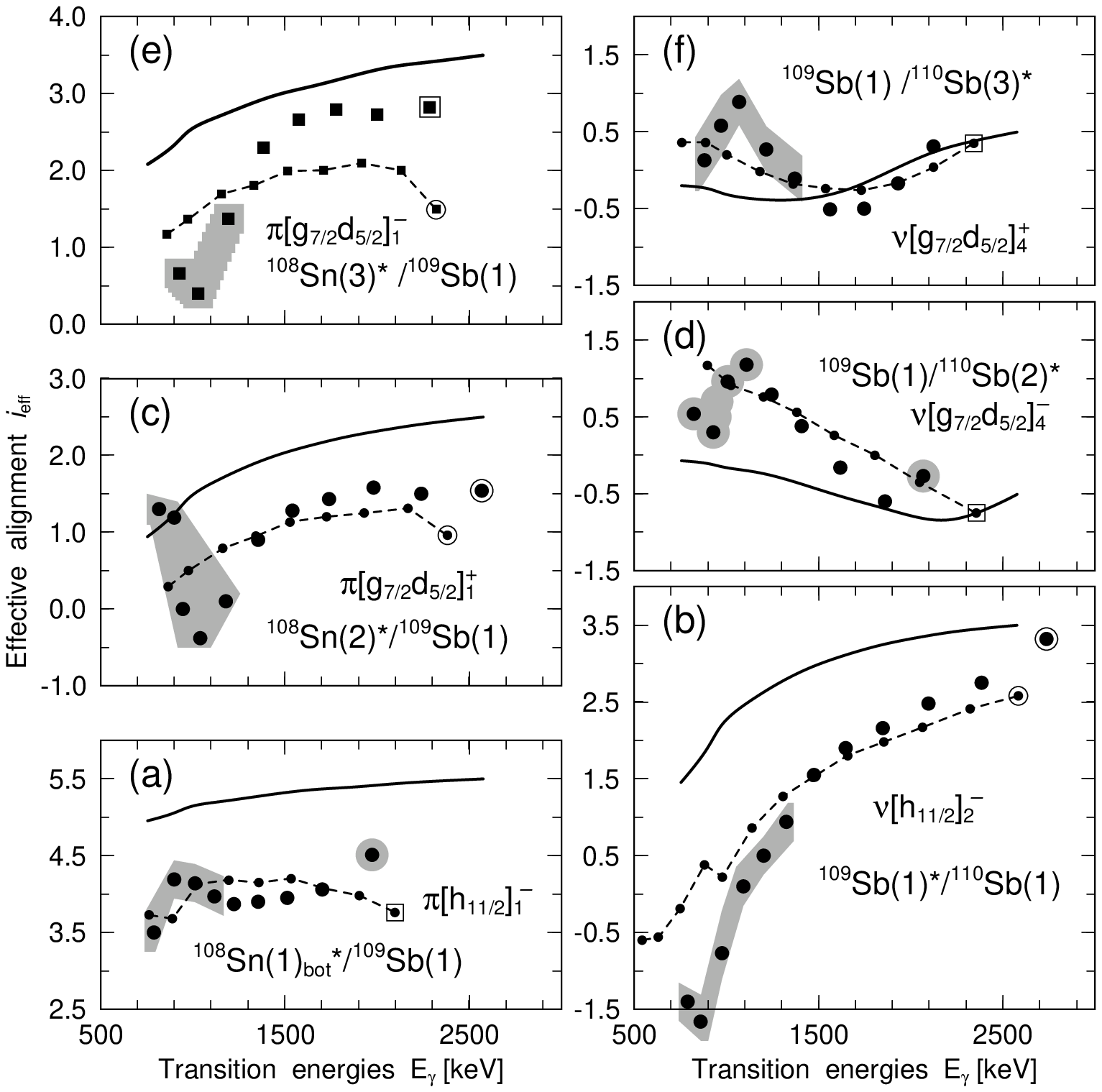}
\vspace*{-2.5cm}
Fig. 8. {\it Similar to Figs. 5 and 6. 
Solid lines are used in order to show the angular 
momentum alignments $<j_x>$ of the single-particle 
orbitals indicated on the panels. The values 
of $<j_x>$ are calculated along the 
deformation path of the $[21,2]^-$ configuration 
in $^{109}$Sb and they are given at the energies
of the $\gamma$-transitions within this 
configuration. 
}
\vspace{0.5cm}

\subsection{Relation between effective alignment $i_{eff}$
and angular momentum alignment $\langle j_x \rangle $ of 
the single-particle orbitals.}

In Fig. 8, the angular momentum alignments $\langle j_x \rangle$ 
of the single-particle orbitals calculated along the deformation
path of the $[21,2]^-$ configuration in $^{109}$Sb are presented. 
They are compared with some corresponding calculated and 
experimental effective alignments. Contrary to the case of the 
SD bands, see discussion in section 2,  the effects 
associated with changes in deformation play a significant
role and the effective alignments of smooth terminating 
bands do reflect not only the alignment properties 
of the single-particle orbital. This is clearly seen from 
the fact that the difference between $i_{eff}$ and 
$\langle j_x \rangle$ reaches $\sim 1.0-1.5\hbar$ for the 
most of the orbitals. It is interesting to note that if the 
absolute $\langle j_x \rangle$ values are large, the differences 
between $\langle j_x \rangle$ and $i_{eff}$ are generally also
large. 

Apart from the case of the $^{108}$Sn(3)/$^{109}$Sb(1) 
pair, the experimental effective alignments agree much 
better with the calculated effective alignments $i_{eff}$ than 
with the pure single-particle alignments $\langle j_x \rangle$. 
This clearly indicates that the changes in equilibrium 
deformation strongly influence the $i_{eff}$ values and 
suggests that these changes are reproduced properly in the 
present approach. As it is discussed in subsection 3.B, the 
less good agreement for the $^{108}$Sn(3)/$^{109}$Sb(1) pair
might be connected with some uncertainty in the 
interpretation of the unconnected band 3 in $^{108}$Sn. 

\section{Conclusions}

Relative properties of smooth terminating bands
observed in $^{108,109}$Sn and $^{109,110}$Sb
nuclei \cite{Sn108,Sn109,Sb109new,Sb110} have 
been studied within the configuration-dependent
shell-correction approach through their effective 
alignments. In the present study, we have used
the configuration and spin assignments given in
original articles for the bands observed in these 
nuclei. It is shown that without any additional 
assumptions, reasonable agreement between theory 
and experiment exists also for effective alignments 
of these bands (with exception of the unconnected band
3 in $^{108}$Sn). This shows that the interpretation 
of smooth terminating bands based on the features of 
experimental and theoretical $(E-E_{RLD})$ curves 
is consistent with the present study. As a result,
the theoretical interpretation using the
parabola-like behaviour of the $(E-E_{RLD})$ 
curves of the bands which terminate in unfavoured 
way appears reliable. Moreover, the present analysis
gives strong support in the spin assignment for
unlinked bands 1-4 in $^{109}$Sb thus indicating 
that they have indeed been observed up to their 
terminating states. 

In addition, our investigation indicates
that the effective alignment approach can be used also 
for the analysis of smooth terminating bands
(and more general for any kind of rotational
bands provided that they are observed over a
considerable spin range) giving an additional 
and very sensitive tool for the interpretation 
of observed bands, especially in the cases
when they are not linked to the low-spin level 
scheme. However, compared with the case of 
superdeformed bands the changes in 
deformation between two bands play a much more important role 
in the case of smooth terminating bands and,
as a consequence, the  effective alignment does 
not necessary come close to the alignment of 
the corresponding single-particle orbital.

We are grateful for financial support from the
Royal Swedish Academy of Sciences, from the Crafoord
Foundation (Lund, Sweden) and from the Swedish Natural
Science Research Council.


\begin{thebibliography}{99}

\bibitem{BR.83} T.\ Bengtsson and I.\ Ragnarsson,
Phys.\ Scripta T 5 (1983) 165

\bibitem{RXBR.86} I.\ Ragnarsson, Z.\ Xing, T.\ Bengtsson and
M.\ A.\ Riley, Phys.\ Scripta 34 (1986) 651

\bibitem{A110} A.\ V.\ Afanasjev and I.\ Ragnarsson,
Nucl.\ Phys. A 591 (1995) 387

\bibitem{158Er} J.\ Simpson, M.\ A.\ Riley, S.\ J.\ Gale,
J.\ F.\ Sharpey-Schafer, M.\ A.\ Bentley, A.\ M.\ Bruce, R.\ Chapman,
 R.\ M.\ Clark, S.\ Clarke, J.\ Copnell, D.\ M.\ Cullen,
P.\ Fallon, A.\ Fitzpatrick, P.\ D.\ Forsyth, S.\ J.\ Freeman, 
P.\ M.\ Jones,
M.\ J.\ Joyce, F.\ Liden, J.\ C.\ Lisle, A.\ O.\ Macchiavelli, 
A.\ G.\ Smith,
J.\ F.\ Smith, J.\ Sweeney, D.\ M.\ Thompson, S.\ Warburton, 
J.\ N.\ Wilson,
T.\ Bengtsson and I.\ Ragnarsson, Phys.\ Lett. B 327 (1994) 187

\bibitem{Sb109} V.\ P.\ Janzen, D.\ R.\ LaFosse, H.\ Schnare, 
D.\ B.\ Fossan, A.\ Galindo-Uribarri, J.\ R.\ Hughes, 
S.\ M.\ Mullins, 
E.\ S.\ Paul, L.\ Persson, S.\ Pilotte, D.\ C.\ Radford, 
I.\ Ragnarsson, 
P.\ Vaska, J.\ C.\ Waddington, R.\ Wadsworth, D.\ Ward, 
J.\ Wilson and 
R.\ Wyss,  Phys.\ Rev.\ Lett. 72 (1994) 1160

\bibitem{Rag95} I.\ Ragnarsson, V.\ P.\ Janzen, D.\ B.\ Fossan,
N.\ C.\ Schmeing and R.\ Wadsworth, Phys.\ Rev.\ Lett. 74 (1995) 3935

\bibitem{Hg194} T.\ L.\ Khoo, M.\ P.\ Carpenter, T.\ Lauritsen,
D.\ Ackermann, I.\ Ahmad, D.\ J.\ Blumenthal, 
S.\ M.\ Fischer, R.\ V.\ F.\
Janssens, D.\ Nisius, E.\ F.\ Moore, A.\ Lopez-Martens, T. D{\o}ssing,
R.\ Kruecken, S.\ J.\ Asztalos, J.\ A.\ Becker, L.\ Bernstein,
R.\ M.\ Clark, M.\ A.\ Deleplanque, R.\ M.\ Diamond, P.\ Fallon,
L.\ P.\ Farris, F.\ Hannachi, E.\ A.\ Henry, A.\ Korichi,
I.\ Y.\ Lee, A.\ O.\ Macchiavelli and F.\ S.\ Stephens,
Phys.\ Rev.\ Lett. 76 (1996) 1583

\bibitem{Pb194} A.\ Lopez-Martens, F.\ Hannachi, A.\ Korichi,
C.\ Sch{\"u}ck, E.\ Gueorguieva, Ch.\ Vieu, B.\ Haas, R.\ Lucas,
A.\ Astier, G.\ Baldsiefen, M.\ Carpenter, G. de France, R.\ Duffait,
L.\ Ducroux, Y.\ Le Coz, Ch.\ Finck, A.\ Gorgen, H.\ H{\"u}bel,
T.\ L.\ Khoo, T.\ Lauritsen, M.\ Meyer, D.\ Pr{\'e}vost, N.\ Redon,
C.\ Rigollet, H.\ Savojols, J.\ F.\ Sharpey-Schafer, O.\ Stezowski,
Ch.\ Theisen, U.\ Van Severen, J.\ P.\ Vivien and A.\ N.\ Wilson,
Phys.\ Lett. B 380 (1996) 18

\bibitem{BRA.88} T.\ Bengtsson, I.\ Ragnarsson and
S.\ {\AA}berg, Phys.\ Lett. B 208 (1988) 39

\bibitem{Rag91} I.\ Ragnarsson, Phys.\ Lett. B 264 (1991) 5

\bibitem{Rag93} I.\ Ragnarsson, Nucl.\ Phys. A 557 (1993) 167c

\bibitem{BR.85} T.\ Bengtsson and I.\ Ragnarsson, 
Nucl.\ Phys. A 436 (1985) 14

\bibitem{Greece} A.\ V.\ Afanasjev and I.\ Ragnarsson,
Proceedings of European Conference on ``Advances in Nuclear 
Physics and Related Areas'', Thessaloniki, Greece, 1997,
in press.

\bibitem{Sn106} R.\ Wadsworth, H.\ R.\ Andrews, C.\ W.\ Beausang, 
R.\ M.\ Clark, J.\ DeGraaf, D.\ B.\ Fossan, A.\ Galindo-Uribarri,  
I.\ M.\ Hibbert,  K.\ Hauschild,   J.\ R.\ Hughes, V.\ P.\ Janzen, 
D.\ R.\ LaFosse, S.\ M.\ Mullins, E.\ S.\ Paul, L.\ Persson,
S.\ Pilotte, D.\ C.\ Radford, H.\ Schnare, P.\ Vaska, D.\ Ward, 
J.\ N.\ Wilson and I.\ Ragnarsson, Phys.\ Rev.\  C 50 (1994) 483

\bibitem{Sn108} R.\ Wadsworth,  C.\ W.\ Beausang, M.\ Cromaz,
J.\ DeGraaf, T.\ E.\ Drake, D.\ B.\ Fossan, S.\ Flibotte,
A.\ Galindo-Uribarri,  K.\ Hauschild, I.\ M.\ Hibbert,
G.\ Hackman, J.\ R.\ Hughes, V.\ P.\ Janzen,
D.\ R.\ LaFosse, S.\ M.\ Mullins, E.\ S.\ Paul,
D.\ C.\ Radford, H.\ Schnare, P.\ Vaska, D.\ Ward,
J.\ N.\ Wilson and I.\ Ragnarsson, Phys.\ Rev. C 53 
(1996) 2763

\bibitem{Sn109} L.\ K{\"a}ubler, H.\ Schnare, D.\ B.\ Fossan,
A.\ V.\ Afanasjev, W.\ Andrejtscheff, R.\ G.\ Allat,
J. de Graaf, H.\ Grawe, I.\ M.\ Hibbert, I.\ Y.\ Lee,
A.\ O.\ Macchiavelli, N.\ O'Brien, K.-H.\ Maier, E.\ S.\ Paul,
H.\ Prade, I.\ Ragnarsson, J.\ Reif, R.\ Schubart,
R.\ Schwengner, I.\ Thorslund, P.\ Vaska, R.\ Wadsworth
and G.\ Winter, Z.\ Phys. A 356 (1996) 235

\bibitem{Sb109new} H.\ Schnare, D.\ R.\ LaFosse, D.\ B.\ Fossan,
J.\ R.\ Hughes, P.\ Vaska, K.\ Hauschild, I.\ M.\ Hibbert, 
R.\ Wadsworth, V.\ P.\ Janzen, D.\ C.\ Radford,
S.\ M.\ Mullins, C.\ W.\ Beausang, E.\ S.\ Paul,
J.\ DeGraaf, I.-Y.\ Lee, A.\ O.\ Macchiavelli,
A.\ V.\ Afanasjev and I.\ Ragnarsson
Phys.\ Rev. C 54 (1996) 1598
 
\bibitem{Sb110}
G.\ J.\ Lane, D.\ B.\ Fossan, I.\ Thorslund, P.\ Vaska,
R.\ G.\ Allatt, E.\ S.\ Paul, L.\ K\"aubler, H.\ Schnare,
I.\ M.\ Hibbert, N.\ O'Brien, R.\ Wadsworth, W.\ Andrejtscheff,
J.\ de Graaf, J.\ Simpson, I.\ Y.\ Lee, A.\ O.\ Macchiavelli,
D.\ J.\ Blumenthal, C.\ N.\ Davids, C.\ J.\ Lister,
D.\ Seweryniak, A.\ V.\ Afanasjev and I.\ Ragnarsson,
Phys. Rev. C 55 (1997) R2127

\bibitem{Te114} I.\ Thorslund, D.\ B.\ Fossan, D.\ R.\ LaFosse,
H.\ Schnare, K.\ Hauschild, I.\ M.\ Hibbert, S.\ M.\ Mullins,
E.\ S.\ Paul, I.\ Ragnarsson, J.\ M.\ Sears, P.\ Vaska and
R.\ Wadsworth, Phys. Rev. C 52 (1995) R2839

\bibitem{Te116}
J.\ M.\ Sears, D.\ B.\ Fossan, I.\ Thorslund, P.\ Vaska, E.\ S.\ Paul,
K.\ Hauschild, I.\ M.\ Hibbert, R.\ Wadsworth, S.\ M.\ Mullins,
A.\ V.\ Afanasjev and I.\ Ragnarsson, Phys. Rev. C 55 (1997) 2290

\bibitem{I113a} M.\ P.\ Waring, E.\ S.\ Paul, C.\ W.\ Beausang, 
R.\ M.\ Clark,
R.\ A.\ Cunningham, T.\ Davinson, S.\ A.\ Forbes, D.\ B.\ Fossan, 
S.\ J.\ Gale,
A.\ Gizon, J.\ Gison, K.\ Hauschild, I.\ M.\ Hibbert, A.\ N.\ James, 
P.\ M.\ Jones, M.\ J.\ Joyce, D.\ R.\ LaFosse, R.\ D.\ Page,
I.\ Ragnarsson, H.\ Schnare, P.\ J.\ Sellin, J.\ Simpson, P.\ Vaska, 
R.\ Wadsworth and P.\ J.\ Woods, Phys.\ Rev. C  51 (1995) 2427

\bibitem{I115b} E.\ S.\ Paul, H.\ R.\ Andrews, V.\ P.\ Janzen, 
D.\ C.\ Radford, D.\ Ward, T.\ E.\ Drake, J.\ DeGraaf, S.\ Pilotte 
and I.\ Ragnarsson, Phys.\ Rev. C 50 (1994) 741

\bibitem{Gd} B.\ Haas, V.\ P.\ Janzen, D.\ Ward,
H.\ R.\ Andrews, D.\ C.\ Radford, D.\ Pr\'{e}vost,
J.\ A.\ Kuehner, A.\ Omar, J.\ C.\ Waddington, T.\ E.\ Drake,
A.\ Galindo-Uribarri, G.\ Zwartz, S.\ Flibotte, P.\ Taras and
I.\ Ragnarsson, Nucl.\ Phys. A 561 (1993) 251

\bibitem{Gd147} Ch.\ Theisen, N.\ Khadiri, J.\ P.\ Vivien, 
I.\ Ragnarsson,
C.\ W.\ Beausang, F.\ A.\ Beck, G.\ Belier, T.\ Byrski,
D.\ Curien, G.\ de France, D.\ Disdier, G.\ Duch{\^{e}}ne, Ch.\ Finck,
S.\ Flibotte, B.\ Gall, B.\ Haas, H.\ Hanine, B.\ Herskind,
B.\ Kharraja, J.\ C.\ Merdinger, A.\ Nourreddine, B.\ M.\ Nyak{\'{o}},
G.\ E.\ Perez, S.\ Pilotte, D.\ Pr\'{e}vost, O.\ Stezowski,
V.\ Rauch, C.\ Rigollet, H.\ Savojols, J.\ Sharpey-Schafer, P.\
J.\ Twin, L.\ Wei and K.\ Zuber, Phys.\ Rev. C 54 (1996) 2910

\bibitem{ALR.97} A.\ V.\ Afanasjev, G.\ Lalazissis and P.\ Ring,
Heavy Ion Physics (Proceedings of International Symposium on 
Exotic Nuclear Shapes, Debrecen, Hungary, 1997), in press and 
to be published

\bibitem{LF.94} D.\ R.\ LaFosse, D.\ B.\ Fossan, J.\ R.\ Hughes, 
Y.\ Liang,
H.\ Schnare, P.\ Vaska, M.\ P.\ Waring, J.-y.\ Zhang, R.\ M.\ Clark,
R.\ Wadsworth, S.\ A.\ Forbes, E.\ S.\ Paul, V.\ P.\ Janzen,
A.\ Galindo-Uribarri, D.\ C.\ Radford, D.\ Ward, S.\ M.\ Mullins,
D.\ Pr\'{e}vost and G.\ Zwartz, Phys.\ Rev.\  C 50 (1994) 1819

\bibitem{RW.93} R.\ Wadsworth, H.\ R.\ Andrews, R.\ M.\ Clark,
D.\ B.\ Fossan, A.\ Galindo-Uribarri,  J.\ R.\ Hughes,   
V.\ P.\ Janzen,  
D.\ R.\ LaFosse, S.\ M.\ Mullins, E.\ S.\ Paul,  D.\ C.\ Radford,
H.\ Schnare,  P.\ Vaska,   D.\ Ward, J.\ N.\ Wilson and R.\ Wyss,
Nucl.\ Phys.\  A 559 (1993) 461

\bibitem{Piaski}
A.\ V.\ Afanasjev, I.\ Ragnarsson and J. M. Sears,
Acta Physica Polonica B 27 (1996) 187

\bibitem{ARunp} A. V. Afanasjev and I. Ragnarsson, 
unpublished


\end{thebibliography}
\end{document}